\documentclass[12pt]{article} 
\usepackage{amsmath} 
\usepackage{amssymb}
\usepackage{epsfig}
\font\openface=msbm10 at10pt
\def\Minkowski     {{\hbox{\openface M}}}

\def\Reals         {{\hbox{\openface R}}}

\def\Hyperbolic    {{\hbox{\openface H}}}

\begin{document}
\title {Quantum Gravity Phenomenology, Lorentz Invariance and Discreteness} 
\author{Fay Dowker\footnote{Blackett Laboratory,
Imperial College, London SW7 2BZ, UK and Perimeter Institute, 
35 King Street North, Waterloo, Ontario N2J 2W9, Canada. E-mail:
fdowker@perimeterinstitute.ca}, 
Joe Henson\footnote{Department of 
Mathematics, University of California/San Diego, La Jolla,
CA 92093-0112, USA. E-mail: jhenson@math.ucsd.edu} 
and Rafael D. Sorkin\footnote{Department of Physics, Syracuse University,
Syracuse NY 13244-1130, USA and Perimeter Institute, 
35 King Street North, Waterloo, Ontario N2J 2W9, Canada. Email: sorkin@physics.syr.edu}} 
\maketitle
\begin{abstract}
Contrary to what is often stated, a fundamental spacetime discreteness
need not contradict Lorentz invariance.  A causal set's discreteness is
in fact locally Lorentz invariant, and we recall the reasons why.  For
illustration, we introduce a phenomenological model of massive particles
propagating in a Minkowski spacetime which arises from an underlying
causal set.  The particles undergo a Lorentz invariant diffusion in
phase space, and we speculate on whether this could have any bearing on
the origin of high energy cosmic rays.
\end{abstract}
\vskip 1cm
%=====================================================

%\section{Introduction}
%\section{Lorentz Invariance and Discreteness}
In discrete approaches to quantum gravity, the fundamental description
of spacetime is not taken to be a manifold, but some discrete structure
to which the manifold is only an approximation.  The scale of this
discreteness is usually assumed to be Planckian.  It is often asserted
any such theory must violate local Lorentz invariance (LLI) and a 
new area of research -- LLI violating phenomenological effects of
quantum gravity -- has grown up around this idea.  The purpose of this
letter is to emphasize that causal set theory \cite{SpacetimeAsCS}
respects LLI and to open a 
new
phenomenological window on this approach to quantum gravity.

What does it mean to say that a discrete theory respects Lorentz invariance? 
It is difficult to give a precise answer, but intuitively the import is
clear.  Whenever a continuum is a good approximation to the underlying
structure (and assuming specifically that the approximating continuum is
a Lorentzian manifold $M$), 
{\it the underlying discreteness must not, in and of itself, suffice to
 distinguish a local Lorentz frame at any point of $M$}.
In consequence, no phenomenological theory in $M$ derived from
such a scheme can involve a local (or global) Lorentz frame
either.\footnote
{Naturally, there can be no question of a literal action of the entire
 Lorentz group on an individual discrete structure.  Rather such a
 structure can only be Lorentz invariant in the same sense
 that a fluid is translation invariant.  This should not detract from
 the fact that a fluid is indeed translation invariant in an important
 sense, whereas a crystalline solid is not.}

Of course the above presupposes an answer to the question: ``How is the
approximating continuum related to the discrete entity that underlies
it?''.  Whether or not a particular discrete theory respects LLI cannot
be settled until this question is answered in the context of that
theory.  Luckily, in causal set theory, there is a clear proposal for
an answer, and we will show that LLI is indeed respected.

A causal set (causet for short) is a locally
finite, partially ordered set (for reviews and motivation for causal
set theory see \cite{Sorkin:1990bh} \cite{Forks}). 
This is a set, $C$, endowed with a binary relation $\prec$ such that 
elements of the set satisfy the conditions 
(i) $(x\prec y)\, {\rm and}\, (y\prec z) \implies (x\prec z)$ (transitivity), 
(ii) $x\not\prec x$ (acyclicity), and 
(iii) all {\it intervals} $\{x: y\prec x\prec z\}$ are finite.  
The relation $\prec$ 
gives rise, in the continuum limit, 
to the causal order on spacetime points, and the
number of elements in a subcauset yields the volume of the
corresponding region of the continuum in Planck units.
 
In the continuum context, the causal
order and volume information suffice to specify a 
(causally reasonable) Lorentzian manifold 
\cite{Hawking:1976fe, Malament:1977}.  
It is therefore reasonable to
regard a Lorentzian manifold as an approximation to a
causet if that causet is a discrete ``sampling'' of the
continuum causal order with uniform density.\footnote
{More generally, one would only require that some coarse-graining of the
 causet approximate $M$ in this sense; but we ignore this distinction here.
 For a somewhat different approach to defining a relation of closeness
 between a manifold and a causet see \cite{Noldus:2003i}.}
More specifically, we may say that 
a Lorentzian manifold $M$ approximates a causet $C$ ($M\approx C$) if
$C$ could have arisen, with relatively high probability, via a random
process of ``sprinkling into $M$'', at Planck density,\footnote
{ By ``Planck density'', we really mean ``density unity in fundamental
 units''.  One expects fundamental units to be equal in order of
 magnitude to Planck units.}
with the causet relations induced by the spacetime causal
structure.\footnote%
{Taking the order relation of the causet to be induced strictly from
 that of the spacetime is only the simplest possibility.  Other rules
 could be considered, but they would not affect anything in this paper.}

A ``sprinkling'' is more properly described as a Poisson process.  To
see what this means,
imagine dividing $M$, using any local coordinate systems, into small
boxes of volume $V$, and then placing a ``sprinkled point''
independently into each box with probability $V/V_{fund}$, where
$V_{fund}$ is the fundamental volume (of order the Planck volume).  The
Poisson process is the limit of this procedure as $V$ tends to zero.
Because spacetime volume is an invariant, the limiting process is
independent of the coordinate systems used to define the boxes.  It
follows that one cannot tell which frame was used to produce the
sprinkling: the approximation is ``equally good in all frames.''

\begin{figure}[ht]
\centering \resizebox{4.4in}{1.1in}{\includegraphics{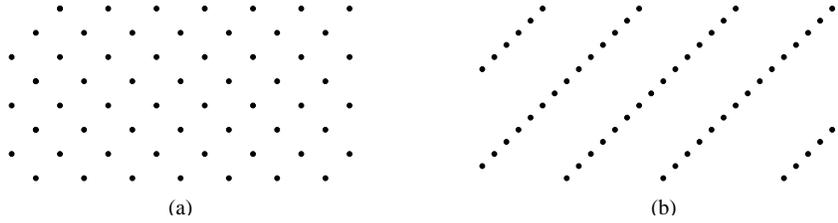}}
\caption{\small{A regular lattice of spacetime points in two different
Lorentz frames.  Normal conventions for spacetime axes are used.
While in (a) the lattice appears to have a regular density of
elements, in the boosted frame shown in (b) the density of points is
revealed not to be uniform.}\label{lattice}}
\end{figure}

Why is the randomness crucial?  
Let us take the example of 1+1 dimensional Minkow\-ski space. 
One obvious way to try to discretize it is
to choose a frame and use a ``diamond lattice'' with respect to that frame,
i.e. the points with coordinates 
$(t,x)=(\epsilon (r+s),\epsilon (r-s))$, 
where $r$ and $s$ are integers and $\epsilon$ is some fixed length,
as in figure (\ref{lattice}a).  
In this frame the lattice 
appears to be a good approximation; all ``nicely shaped'' large
regions have a similar density of elements.  In a frame boosted at
velocity $V$ in the positive direction, however, the elements are at
$(\gamma \epsilon [(1-V)r+(1+V)s], \gamma \epsilon[(1-V)r-(1+V)s])$.
Figure (\ref{lattice}b) shows this lattice with $\gamma=1.25$.  
Now it becomes clear that, 
if the boost is large enough, there will be
``nice'' big regions containing no elements at all,
and others containing far too many elements. 
The approximation only looks good in the original frame, and so it
breaks Lorentz invariance by preferring this frame.
In light of this example, it seems likely
that the same problem would affect any non-random discretization 
of a spacetime.  
Thus, for example, 
a Regge-type triangulation whose simplices look ``fat'' in one frame
will look ``long and skinny'' in a relatively highly boosted frame.

We want to emphasise that not only is the {\it process} of 
sprinkling Lorentz invariant but so 
also 
are almost all of the individual causets that are generated.  
An objection that often comes up in this connection concerns the
necessary occurrence of voids in any 
given 
Poisson sprinkling.  
While it is true that voids must occur, 
this does not cause a problem for
Lorentz invariance (or any other problem that we know of).  
However, one may still feel uneasy about the voids, and some people seem
to believe that they necessarily {\it would} break Lorentz invariance
in some manner.
To put such
qualms in perspective, let us estimate the probability that there is at
least one void of nuclear dimensions in the history of the observable
universe since the Big Bang.  More precisely, we will bound the
probability that a sprinkling would leave empty any interval whose
height is of the order of one Fermi.  (An interval in spacetime is a
``double light cone'' or ``Alexandrov neighborhood''.  We do not require
the ``axis'' of the interval to be aligned with the cosmic rest frame.
Hence our bound will apply to the probability of ``finding a void in any
frame''.)

All the numbers in what follows
are ``of the order of''.  
Consider as a model of the universe a portion $P$ of 
Minkowski spacetime, the size of the observable universe
and defined by $0\le t\le T$, $0 \le x^i \le T$, 
$i = 1,2,3$ in some frame. 
If $T$ is 13 billion years, the spacetime 
volume is $10^{240}$ in fundamental units.
An interval of nuclear size has spacetime volume 
$10^{80}$. 
This means that the probability that any {\it particular}
nuclear sized interval will be a void is $e^{-10^{80}}$.
But we want the 
probability $q$ that at least one interval (any one) in $P$ will be void. 
This can't be calculated easily, because the intervals overlap and
the probabilities for them to be void are not independent.  However, we
can put an upper bound on $q$ without much difficulty.

Let us fill $P$ with coordinate balls  (with respect
to the ``defining frame'' of $P$) of small enough radius  
that any ``upright'' interval of nuclear size is guaranteed to contain 
at least one complete ball.  
(Upright means that the top and bottom points have the same spatial
coordinates.)
A radius of one hundredth nuclear size will do,
and we will have $10^{168}$ of these balls packed into $P$. 
The probability that at least one of them is void is $10^{168} e^{-10^{72}}$.
These same balls will also suffice for intervals of  
nuclear volume that are slightly boosted or ``tilted'' from the upright. 
They will certainly do for all $\gamma$ factors less than or equal to 
$\gamma = 5/4$, that is for a region of the Lorentz group with volume 
of order 1.  For each such cell of the Lorentz group we choose
a set of coordinate balls in spacetime (in the corresponding  frame).
The relevant region of the Lorentz group is bounded by the maximum 
relevant boost in $P$, which corresponds to a $\gamma$ factor of $10^{42}$.
(Any larger boost would produce a ball that could not fit into $P$.
In fact the maximum $\gamma$ is actually smaller, since the boosted nucleus
would meet the boundary of $P$ before the smaller ball would.)
The number of cells needed to cover this region of the Lorentz group is
$10^{84}$.  
The probability of getting any nuclear sized void is less than the
probability that any one of the coordinate balls from any of the boosted
sets will be void.  This in turn is less than
$10^{84}\times{10}^{168}{\times}e^{-10^{72}}$, a number so tiny that
the two prefactors have no impact whatsoever on its value.

Given that causal sets respect Lorentz invariance, what conclusions can
be drawn?
Most obviously, we predict that no violation of LLI will be observed at
the phenomenological level, so that, if any of the experiments currently
planned or under way did find such a violation, the causal set hypothesis
would be disfavored.
But this is only a negative prediction.  Are there also positive signatures?
Since 
the causal set hypothesis makes such a definite 
statement about the underlying structure for spacetime and how it 
is related to the continuum we actually experience, 
it is not difficult to devise concrete models predicting 
potentially observable effects of the underlying discreteness.
We give one such model below which we call ``swerving'', after Lucretius 
\cite{Lucretius}:

``The atoms must a little swerve at times -- 
but only the least, lest we should seem to feign 
motions oblique, and fact refute us there.''

In the continuum, massive particles travel on timelike geodesics.  
However,
an underlying discreteness might induce small
fluctuations in the particle's worldline, and the causal set picture
naturally suggests models in which this effect would be Lorentz
invariant.  Though it might be too small to observe on everyday scales,
such an effect might be detectable in sensitive laboratory experiments,
or by astronomical observations if the particle were travelling over
cosmic distances.  Here is a model of this type.

Consider a hypothetical point-particle 
of mass $m$
moving through a causet $C$
derived by sprinkling Minkowski space $\Minkowski^4$.  We will take its
trajectory to be a {\it chain} of elements of $C$ (i.e. a totally
ordered subset of $C$).  No such chain can correspond perfectly to a
straight line in $\Minkowski^4$.  So, given an ``initial segment'' of
the trajectory up to some element $e_n$ of $C$, how could its future
continuation be determined?  
If we assume that the trajectory's past determines its future, but that
only a certain proper time $\tau_f$ into the past of the trajectory is
relevant, then we are led to the following Lorentz invariant rule as
a particularly simple discrete analog of geodesic motion.

\begin{figure}[ht]
\centering \resizebox{3in}{3in}{\includegraphics{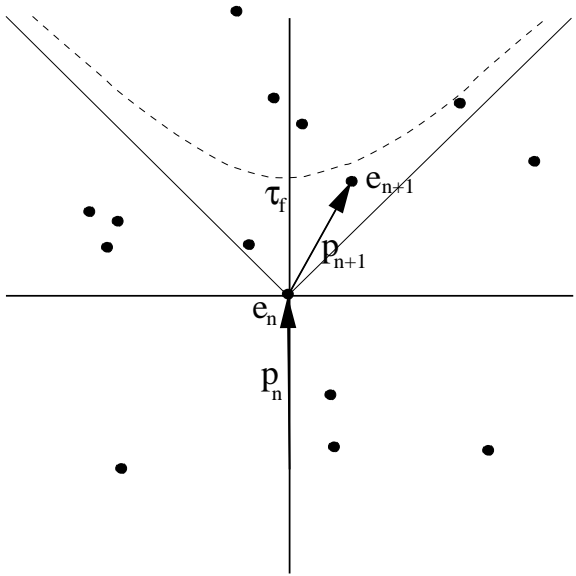}}
\caption{\small{A portion of $1+1$ Minkowski space where the dots
represent elements of a causet sprinkled into it.  The trajectory of
the particle has reached $e_n$ with momentum ${p}_n$ (the frame
having been chosen so that the three-momentum is zero). The dotted
line is the hyperbola of points a proper time $\tau_f$ to the future of
$e_n$.  The element within proper time $\tau_f$ of $e_n$ that best
preserves the momentum is $e_{n+1}$.  The ratio of $\tau_f^{-2}$ to the
density of sprinkling has been exaggerated here to emphasize the
momentum change in one step.  In a more realistic model $\tau_f$ would
be larger.}\label{sprinkle}}
\end{figure}

The future trajectory is constructed inductively, as illustrated in
fig. \ref{sprinkle}.  Starting from an element $e_n$ and a momentum
${p}_n$, the next element in the trajectory, $e_{n+1}$, must be chosen.
For convenience we have drawn the diagram in the rest frame at $e_n$.  
The new momentum ${p}_{n+1}$ is defined to be
proportional to the vector between $e_n$ and $e_{n+1}$.  The selection
of $e_{n+1}$ is made such that $e_{n+1}$ is in the causal future of
$e_n$ and is within proper time $\tau_f$ of $e_n$ and so that
$|{p}_{n+1}-{p}_n|$ is minimized.  In the figure, this means that
$e_{n+1}$ is within the future light cone of $e_n$, below the dotted
hyperbola, and such that the vector from $e_n$ to $e_{n+1}$ is as close
to the vertical as possible. (That there exists such an element in that
region is guaranteed by the infinite spacetime volume of the region, the
Poisson distribution and the local finiteness condition.)  This realizes
the ideas that the trajectory should be as close to a straight line as
possible and that the dynamics should be approximately Markovian if the
``forgetting time'' $\tau_f$ is small.  (There is ample room for it to
be small and yet much bigger than the discreteness scale if the latter
is Planckian.)  The process is then repeated starting with $e_{n+1}$ and
${p}_{n+1}$.

Our model implies random fluctuations in the momentum of the particle. 
For any numbers $\delta_1 > \delta_2 > 0$
at stage $n$
there is a finite volume within proper time
$\tau_f$ to the future of $e_n$ such that, if element 
$e_{n+1}$ were in this volume, the momentum change 
$|{p}_{n+1}-{p}_n|$ would lie between $\delta_1 $ and $\delta_2$.
The probability of this happening is the probability that 
this volume is not empty of sprinkled points,
while the volume leading to a smaller momentum change is empty.
Both these probabilities are given precisely by the Poisson distribution: 
the probability that a volume $U$ is empty is $e^{-\varrho U}$, 
where $\varrho$ is the density of the sprinkling. 

This simple model can be criticised on many grounds: it treats the particle
as if it were of zero size, it is deterministic rather than quantum,
{\textit etc.}.  
Most seriously, it could not possibly be fundamental, since the law of
motion of the trajectory is not formulated in terms of the causet, but
refers also to the approximating Minkowski spacetime.\footnote
{This defect can be overcome fairly easily, however.}
We present it 
in the spirit of \cite{Amelino-Camelia:2002vw}: form a concrete model 
with testable consequences based on important aspects of the fundamental
formalism and compare to observation.

Since the hypothesized Lucretius effect is supposed to occur on very
small scales, it should be possible to approximate it over macroscopic
distances by a diffusion equation (hydrodynamic limit).
This is analogous to how the ordinary diffusion equation
describes the long time behaviour of a random walk.
In our case however, the
diffusion is not in physical space $\Reals^3$, 
but on the phase space $\Hyperbolic^3\times\Minkowski^4$,
where $\Hyperbolic^3$ is the mass shell (Lobachevskii space), 
$\Minkowski^4$ is Minkowski spacetime, and 
the diffusion takes place in {\it proper time} $\tau$. 
The diffusion in spacetime is
secondary and is driven by that in momentum space, in close analogy with
the Ornstein-Uhlenbeck process. 

Consider the scalar (not scalar density)
probability distribution $\rho\equiv\rho(p^\nu, x^\mu; \tau)$
on $\Hyperbolic^3\times\Minkowski^4$.  
It is a function of momentum, $p^\nu$, 
spacetime position, $x^\mu$, and proper time
$\tau$. 
We write the full four-momentum 
as an argument with the understanding that on the mass shell there are
only three independent components.   
With the condition 
that the process be Lorentz
invariant, 
the following equation for $\rho$ can be derived by following the
prescription set out in \cite{Sorkin:1986} for stochastic evolution 
on a manifold of states:
\begin{equation}
\label{ptswerves}
  \frac{\partial\rho}{\partial\tau} 
  = k \nabla_{p}^2 \rho 
  - \frac{1}{mc^2} p^\mu \frac{\partial}{\partial x^\mu} \rho
\end{equation}
where $\nabla_{p}^2$ is the Laplacian on
$\Hyperbolic^3$, $m$ is the mass of the particle, 
and $k$ is a constant (that will depend on the parameters of the
discrete process, such as the forgetting time $\tau_f$).

This equation defines a diffusion process in 
which the particle's proper time $\tau$ serves as time.
% parameter.
It is the unique 
Markovian, 
Poincar{\'e} invariant, 
relativistically causal,   
diffusion law in vacuum that preserves $\rho\ge0$.\footnote
{The Ornstein-Uhlenbeck process for a particle diffusing in interaction
 with a relativistic fluid has been considered in \cite{Debbasch:1998}.
 In that case the rest frame of the fluid provides a preferred frame for
 the stochastic noise term driving the diffusion.  The present process
 has no preferred frame and is fundamentally Lorentz invariant.}
For fixed mass, 
it has a single free parameter, the diffusion constant $k$.
(In principle, the coefficient of the 
$p^\mu\frac{\partial}{\partial x^\mu} \rho$ term could be different, but 
that would mean that $p^\mu$ would not be the physical momentum.) 
Given these uniqueness properties, the equation should be insensitive to 
variations in the microscopic model that underlies it, so 
long as the latter is Lorentz invariant, causal and (approximately) Markovian. 
So although the above discrete swerve model might be wrong in 
detail, 
the macroscopic phenomenology of ({\ref{ptswerves}}) transcends it.

If $\rho$ is a function of momentum alone 
and is initially (at $\tau=0$)
a delta function in momentum,  
then the 
(un-normalized) solution -- adapted from the
solution of the diffusion equation 
on $S^3$ \cite{Perrin:1928} -- is 
\begin{equation}
\label{solution}
  \rho(p) = 
  e^{-R^2/4\hat{k}\tau} (\hat{k}\tau)^{-\frac{3}{2}} 
e^{-\hat{k}\tau} \frac{R}{\sinh{R}}
\end{equation}
where $\hat{k} = k/m^2c^2$,  $R= \sinh^{-1}({p/mc})$,
and $p \equiv |{\bf p}|$ is the norm of the three-momentum in the frame 
defined by the point in $\Hyperbolic^3$ at which the diffusion begins.
$\sigma = mc R$ is then the geodesic distance from that point.

Equation (\ref{ptswerves}) and its
fundamental solution ({\ref{solution}}), 
expressed as they are in terms of proper time, 
are not well suited to comparison with experiment/observation,
even though they exhibit the underlying Lorentz invariance very clearly.
Instead, one needs a description of the same process with respect to
cosmic or laboratory time.
To this end, fix a preferred set of spacelike hypersurfaces $t\equiv
x^0=$ constant, and assume that the distribution on some initial
hypersurface of the set is uniform in space
({\it i.e.} $\rho$ does not depend on $x^i$, where $i$ are spatial
coordinates).
The swerves, being spatially homogeneous and isotropic, 
will preserve this uniformity, and so if we assume that any additional 
frictional effects are also homogeneous and isotropic, 
the distribution will remain uniform. 
We require, under these conditions, an equation governing  
how the probability distribution evolves in cosmic time $t$.

Such an equation can be deduced from (\ref{ptswerves}).
(Details of the derivation and of the inhomogeneous case will appear
elsewhere.) 
The result is:
\begin{equation}
\label{swerves}
    {\frac{\partial\rho}{\partial t}} 
%%    {\partial\rho\over\partial t} 
    = k \nabla^2_{p} \left(\frac{\rho}{\sqrt{1+p^2/m^2}} \right)  
    - \nabla_a (w^a \rho)
\end{equation}
Here, the scalar function $\rho\equiv\rho(p^\mu; t)$ on $\Hyperbolic^3$, 
gives the momentum distribution,
$\nabla_a$ is the covariant derivative on $\Hyperbolic^3$,
and $p$ is the norm of the particle's three-momentum,
${\bf p}$ in the cosmic frame. 
(To make this equation 
plausible, notice that the factor 
 $\sqrt{1+p^2/m^2}$ 
 is the boost factor $\gamma=dt/d\tau$.)
The term involving the vector $w^a$ is a 
friction term added to represent the effect on the particle's momentum
of, for example, the Hubble expansion and interactions with the CMBR
(cosmic microwave background radiation).
The specific form of $w$, 
which will in general be a function of the momentum, 
will depend on the type of friction involved.\footnote
{In the case of violent momentum transfers, this would have to be
 generalized to a Boltzmann type collision term.}
For the cases mentioned,  
$w^a$ will have only a radial component 
(in the $p \equiv |{\bf p}|$ direction).

In the time-dependent case, 
(\ref{swerves}) will probably have to be solved numerically. 
However, we can hope to analyze its equilibrium solutions analytically.
For example, 
at large $p$, energy $E\sim p$,
and if $w^p\sim -bE^n$ with $n\ge 1$, 
then the equilibrium solution will behave as $Ee^{-b E^n/n}$. 
If the dominant friction over any high-energy range is constant,
{\it i.e.} 
if $w^p\sim dE/dt \sim$ constant,
then the equilibrium distribution will be a power 
law in that range.

How might one observe diffusion of the above sort?
Cosmological and astrophysical observations are the obvious places
to look for consequences of a universal acceleration mechanism.  
But first, laboratory physics can put an upper bound on the diffusion
constant $k$. 
Suppose the particles in question are protons.
If $k$ were large enough, 
hydrogen gas would spontaneously heat up in a
short time, and this has not been observed.  
In the laboratory regime, 
hydrogen is non-relativistic,
so equation (\ref{swerves}) can be approximated as
\begin{equation} 
   \frac{\partial \rho}{\partial t}= k \nabla^2 \rho
\end{equation}
where $\nabla^2$ is now the standard Laplacian on $\Reals^3$.  This is the
standard diffusion equation and has the well known solution:
\begin{equation}
   \rho = A(t) \exp(-\frac{{p}^2}{4kt})
\end{equation}
where $A(t)$ is a normalization factor.  
Usefully, this is also the form of the Maxwell distribution for 
a classical gas in thermal equilibrium: 
\begin{equation}
   \rho_{\rm Maxwell} = A \exp(-\frac{{p}^2}{2mk_B T})
\end{equation}
where $m$ is the  molecular mass,
$T$ is the temperature,  
and $k_B$ is Boltzmann's constant.  
If a gas starts in a thermal state, 
it will therefore remain in a thermal state even if swerves are included.  
Moreover, the above two equations imply that the temperature will scale
linearly with time, specifically:
\begin{equation}
  \frac{dT}{dt}=\frac{2k}{mk_B}
\end{equation}
Assuming, for the sake of argument,
that a heating rate of a millionth of a degree per second would already
have been detected in the laboratory,  
we obtain the approximate bound
\begin{equation}
   k \leq 10^{-56} kg^2 m^2 s^{-3}
\end{equation} 
The maximum average energy gain due to swerves consistent
with this rate of temperature gain can be obtained from the formula
$\langle E \rangle=3 k_B T/2$: 
\begin{equation}
  \label{DeltaE}
  \langle\Delta E\rangle / \Delta t
  \leq 4.3 \times 10^{-11} {\rm eV s}^{-1}
\end{equation} 
% NB: 6.9x10^{-30} j s{-1}

Now, let us turn to some possible astrophysical effects of swerves. 
One outstanding astronomical puzzle
is
the origin of high energy cosmic rays
(see \cite{Anchordoqui:2002hs} for a recent review). 
Attention is often focused on the so-called ``trans-GKZ'' events, 
apparent detections of cosmic rays with energies above $5\times 10^{19}$ eV. 
Such primaries, if they are protons, 
cannot have come from farther than about 20 Mpc 
(because they would have decayed due to photo-pion production with
 the CMB photons),
but they have no obvious source in that distance range. 
But even for 
cosmic rays between $10^{15}$ eV and $10^{19}$ eV,
there are only suggestions and no universally accepted 
acceleration mechanism for producing the observed energy distribution. 
The data 
(see e.g. fig 1 of \cite{Anchordoqui:2002hs}) 
seem to cry out for a universal cosmic
acceleration mechanism that would inject protons, say, into 
the galaxy with a power law distribution of $E^{-a}$, 
where $2<a<3$, 
so that the observed variations in the power law and
deviations from isotropy would be due to the 
dynamics of the protons in the galaxy.
Could swerves provide such a cosmic mechanism?\footnote
{The idea that the rays might be the result of spontaneous
acceleration, as a result of non-standard QFT, has been discussed
in \cite{Rueda:1999dy}.}

Swerves induce a ``statistical acceleration'' analogous to Fermi
acceleration, and it is possible a priori that enough  
intergalactic hydrogen could be accelerated up to very high energies to
explain the data.
Unfortunately this degree of acceleration is inconsistent with the bound
on $k$ already discussed. 
A ``Lucretian'' explanation of the cosmic ray data, 
assuming the primaries are protons, 
would require some protons to accelerate
to $\sim10^{20}$ eV from far lower energies on a timescale of 
the age of the universe. 
To produce a power law distribution in energy, 
a significant proportion of those protons
reaching, say, $10^{18}$ eV would have to go on to double 
their energy and more.  
The rest frame of a proton with an energy of $10^{18}$ eV
has a $\gamma$ factor of $10^9$ relative to the cosmic frame, 
so 10 billion years of cosmic time is only 10 years of proper time for such
a proton.  
Doubling its energy in the cosmic frame would mean gaining
around 250 MeV in its own frame.
But from
the inequality (\ref{DeltaE}), 
in this frame the average energy gain in 10 years could be at most 
$1.4 \times 10^{-2}$ eV.  
(At these energies, we can trust the non-relativistic approximation.)  
Since the distribution of momentum is Gaussian, 
with such a low average energy gain, 
the probability of gaining 250 MeV is exponentially small.  
In other words, 
a proton has practically no chance of making it from
$10^{18}$ eV to $2\times10^{18}$ eV in the age of the universe as a
result of swerves.  
This calculation assumes that $k$ is roughly 
the same for an intergalactic hydrogen atom as it is for a proton, as
it is for a ${\rm H}_2$ molecule in a box
of gas, 
but the argument is still valid even for a $k$ many orders of
magnitude larger than has been assumed.

Proton swerves cannot explain trans-GKZ cosmic rays either
(if indeed their apparent observation turns out to be correct).
Swerves would accelerate some protons 
back up beyond $10^{20}$ eV 
after they entered the ``GKZ sphere'', $\sim 20$ Mpc from us;
but this effect would not be significant.
The argument for this is similar to that above.  
A proton with an energy of $10^{19}$ eV would reach us from the GKZ
sphere in about 2 hours of proper time.
This would not leave enough time for a non-negligible fraction of
protons to, say, double their energy.

So the most direct application of the swerve idea to 
protons cannot explain the origin of high energy cosmic rays.
However, more complicated scenarios can be considered.
For example, in \cite{Yoshida:1998} the authors postulate 
homogeneously distributed sources producing (by some unknown mechanism) 
neutrinos with energies above $10^{22}$ eV 
which collide with a cosmological background of neutrinos (hot dark matter) 
to produce -- amongst other things -- 
protons and gamma rays 
that could be cosmic ray primaries. 
Perhaps swerves could provide 
the required acceleration in this case. 
Indeed neutrinos are more likely candidates 
than protons to be affected by the underlying discreteness as they
are more point-like than protons, according to present beliefs.
Moreover, we have few if any laboratory bounds on $k$ to contend with in
the case of neutrinos.

More sophisticated models could also be developed. 
Our simple proton model assumes that
the ``diffusion constant'' $k$ 
does not depend on local factors like
average particle density, temperature etc.  
In a more realistic swerve model, 
perhaps a high particle density would lower the rate of diffusion, 
in which case the constraints from laboratory physics could be loosened.
A second sort of generalization of the diffusion (or Fokker-Planck)
equation ({\ref{swerves}}) would relax the assumption of locality in
{\it momentum} space. 
The ``friction term'' ({\ref{swerves}}) 
captures the effect of many small
momentum transfers due to (for example) CMBR scattering. 
The effect of large kicks would have to be described by a Boltzmann equation. 
A third improvement to our model would be to treat the particles 
quantum mechanically rather than classically, allowing one to
take into account
the finite size of the ``wave packet''. 
Such a change might be
important, because it is conceivable that matter-induced decoherence
would influence the value of $k$, making it different on earth than 
in interstellar space.  Unfortunately, however, the type of over-arching
framework that is available for classical diffusion seems to be lacking
in the quantum case, so it is less obvious how to proceed.
Finally, in a full theory of causal set quantum gravity, regions of
continuum spacetime might be best described as a quantum superposition
of many causets, and a better phenomenological model might have to
reflect this aspect as well.

Let us return for a moment to possible observational evidence for
swerves.  
Since the path of a particle would no longer be an exact
geodesic, a certain amount of fuzzing of distant sources of particles
would occur.
Perhaps this could be revealed by highly directional detectors of some
sort.

So far we have limited our discussion to the case of massive particles.
A Lorentz invariant diffusion equation for massless particles can also
be written down, 
although in this case we lack a concrete model of 
propagation on the underlying causet that could serve as motivation.~\footnote
{We also lack an idea of how to describe such a diffusion in wave
 language as opposed to particle language.  What would it mean in
 the case of Maxwell's equations for example?}
Rather than diffusion in $\Hyperbolic^3$, 
we have in this case diffusion on a ``light'' cone, 
since the 4-momentum of a massless particle is a null vector.  
We will describe this further in 
a 
future work.
Lorentz invariant diffusion on the cone cannot alter the direction of
the momentum, but it will cause its magnitude to fluctuate,
so that a distribution peaked at 
a certain energy would spread with time.
Accordingly, 
one might seek evidence for this kind of diffusion in the blurring of
sharply peaked spectral features of distant sources
(such as emission and absorption lines).
This is in contrast to what has been proposed in 
Lorentz-violating models, where the speed of a photon is  
presumed to vary with its energy
(see e.g. \cite{Jacobson:2003bn}). 

Finally, just as tests of Lorentz invariance push special
relativity to its limits, tests of unitarity would 
allow one to
push the
state-vector formalism of quantum mechanics to {\textit{its}} limits.
It seems only reasonable therefore that simple models of the possible
effects of non-unitarity be formulated.  A non-relativistic example 
is given in \cite{Collett:2002}.  Perhaps a relativistic version 
of this could produce a similar effect to swerves.

To summarize, there is no reason that the assumption of an underlying
spacetime discreteness 
must give rise to violations of local Lorentz invariance, because the
causal set hypothesis does not.
To illustrate the
point that discreteness can nevertheless have observable effects, 
we have exhibited a Lorentz invariant momentum 
diffusion motivated by causal sets. 
If it is indeed the case
that certain proposals for quantum gravity, 
such as loop quantum gravity and spin foams 
do predict violations of LLI, then we are in the
happy situation of having a way to distinguish between
different proposals experimentally. 
Be that as it may, 
the specificity of the models treated in this paper indicates
that the causal set approach holds great
potential for providing phenomenological theories of matter propagating 
in a discrete background. 
In an era of ever increasing sensitivity and
power in cosmological observations, 
this potential to predict and detect the effects of 
a fundamental spacetime discreteness should be exploited.

{\bf Note added:}  
The Poincar{\'e}-invariant diffusion process described above
is constructed rigorously in \cite{Dudley:1965} and \cite{Dudley:1967}.

We are grateful to Pasquale Blasi, David Craig,
Stuart Dowker,
Seth Major, Laura Mersini, Angela Olinto
and David Rideout for discussing these ideas with us. 

This research was partly supported 
by NSF grant PHY-0098488 at SU, 
by Air Force grant AFOSR at UCSD,
by the Office of Research and Computing of Syracuse University
and by the Department of Physics at Queen Mary, University of London. 

\bibliographystyle{JHEP} 
  % \bibliographystyle{h-physrev3} 
  % this works: \bibliographystyle{ieeetr}   % -- DPR'sfavorite
  % this works: \bibliographystyle{plain}

\bibliography{first}

\end{document}